\documentclass[conference,10pt]{IEEEtran}
\makeatletter
\def\ps@headings{%
\def\@oddhead{\mbox{}\scriptsize\rightmark \hfil \thepage}%
\def\@evenhead{\scriptsize\thepage \hfil \leftmark\mbox{}}%
\def\@oddfoot{}%
\def\@evenfoot{}}
\makeatother
\usepackage{url,comment,cite}
\usepackage{amsmath,amssymb,amsfonts,bbm,dsfont}
\usepackage{graphicx}
\usepackage{epsfig,epsf,psfrag,textcomp,tabularx}
\usepackage[usenames,dvipsnames]{xcolor}
\usepackage{subfigure}
\usepackage{float}
\usepackage{stfloats,mathtools}
\usepackage[algoruled,medskip,dontprintsemicolon,linesnumbered,Algorithm]{algorithm}
\usepackage[noend]{algpseudocode}
\usepackage{amsmath,mathtools}

\newcommand{\Bc}{\mathcal{B}}
\newcommand{\Cc}{\mathcal{C}}
\newcommand{\Dc}{\mathcal{D}}
\newcommand{\Kc}{\mathcal{K}}

\newcommand{\Rc}{\mathcal{R}}

\newcommand{\Uc}{\mathcal{U}}
\newcommand{\Ab}{\mathbf{A}}
\newcommand{\Cb}{\mathbf{C}}
\newcommand{\Vb}{\mathbf{V}}

\newcommand{\av}{\mathbf{a}}
\newcommand{\sv}{\mathbf{s}}

\numberwithin{equation}{section}

\begin{document}

\title{Fast Resource Scheduling \\
in HetNets with D2D Support}


\author{\IEEEauthorblockN{Francesco Malandrino}
\IEEEauthorblockA{
Politecnico di Torino\\
Torino, Italy \vspace*{-12pt}}
\and
\IEEEauthorblockN{Claudio Casetti}
\IEEEauthorblockA{
Politecnico di Torino\\
Torino, Italy \vspace*{-12pt}}
\and
\IEEEauthorblockN{Carla-Fabiana Chiasserini}
\IEEEauthorblockA{
Politecnico di Torino\\
Torino, Italy \vspace*{-12pt}}
\and 
\IEEEauthorblockN{Zana Limani}
\IEEEauthorblockA{
Politecnico di Torino\\
Torino, Italy \vspace*{-12pt}}
}

\maketitle
\thispagestyle{empty}\pagestyle{plain}

\begin{abstract}
Resource allocation in LTE networks is known to be an 
NP-hard problem. In this paper, we address an even more complex
scenario: an LTE-based,  2-tier heterogeneous network 
where D2D mode is supported under the network control. 
All communications (macrocell, microcell and D2D-based) share
the same frequency bands, hence they may interfere.
We then determine (i) the
network node that should serve each user and (ii) 
the radio resources to be scheduled for such communication.
To this end, 
we develop an accurate model of the system and apply approximate
dynamic programming to solve it. 
Our algorithms allow us to deal with realistic,
large-scale scenarios. In such scenarios,  
we compare our approach to  today's networks where eICIC techniques 
and proportional fairness scheduling are implemented.
Results highlight that our solution increases the system throughput  
while greatly reducing energy consumption. We also show that D2D mode can
effectively support content delivery without significantly harming 
macrocells or microcells traffic, leading to an increased system capacity.
Interestingly, we find that D2D mode can be a low-cost alternative
to microcells. 
\end{abstract}

\section{Introduction\label{sec:intro}}

The deployment of Heterogeneous Networks
(HetNets)~\cite{paradigm-shift}
is a cost-effective solution to ever-growing traffic demands.
Heterogeneity in their design is achieved through 
a multi-tier architecture, i.e., 
a mix of macrocells and smaller cells, namely microcells, picocells
and femtocells.
The benefits of spatial spectrum reuse, which comes with the 
proximity between access network and end users, amply justify
the new technical challenges.
Among such challenges is the likelihood of cross-tier interference 
brought about by intense frequency reuse in neighboring or 
overlapping cells. 
Techniques to mitigate such  superposition of transmission resources
are already available, e.g., ICIC (Inter-Cell Interference Coordination).

However, innovations and challenges introduced 
by the heterogeneity of future networks 
does not stop at cell coverage. 
As a further solution to improve spectrum
utilization, 
User Equipments (UEs) are expected to be able to communicate 
in a device-to-device (D2D)
fashion~\cite{underlay-twc,zulhasnine}.
Such D2D links
will be established 
on LTE licensed bands, as foreseen 
by the 3GPP ProSe group working on Release 12~\cite{3gpp-r12}.
This communication paradigm (commonly referred to as in-band underlay D2D)
will likely be implemented under the control of the cellular infrastructure 
(e.g., Base Stations, BSs)~\cite{shen-opcontrol}. In D2D mode, a UE 
(called serving UE) can forward
to another UE content it has previously downloaded from a network node. 
However, the presence of a serving UE in a specific area is ephemeral, forcing resource 
allocation procedures to promptly adapt to changes in the availability of such nodes.

In our paper, we address the challenges above by proposing a model for
heterogeneous, LTE-based networks. We assume 
that radio resources in such a network are managed by an area controller,
which forwards its decisions to BSs, using a high-speed link
\cite{freescale,ericsson}. 
Such a scenario accounts for the coexistence and integration between 
I2D (Infrastructure-to-Device) 
and network-controlled D2D communication paradigms. 
Under this framework, we answer the following questions: (i) which network node 
(macrocell BS, microcell BS, UE) should serve a UE and (ii) which 
radio resources should be used? 
Answers to these questions will aim at
reducing interference owing to spatial reuse of radio resources,
hence ensuring higher data rates. As a side effect, for a fixed amount
of transferred data, this will also lead to a significant reduction in
the system  energy consumption. 

Resource allocation in LTE is performed on a short time period (1~ms) basis.
We therefore develop a system model using dynamic programming, which is 
particularly suitable to update decisions every time period.
Then, since the resource allocation problem in (even simpler) LTE scenarios is known to be 
NP-hard~\cite{zulhasnine,noi-mascots13}, 
we apply Approximate Dynamic Programming (ADP) to solve the model.
We remark that our ADP methodology
yields a very efficient solution strategy, which
caters for the swiftness required by 
real-world LTE scenarios. 
We  compare our solution to a scenario representing 
today's networks, where 
standard eICIC (enhanced ICIC) techniques are implemented and 
proportional fairness is used for
traffic scheduling at BSs. Results highlight that the ADP approach 
combines energy-efficiency with an increased throughput and it fully
exploits the potentiality of D2D transfer. 
Additionally, thanks to the limited interference when compared to the I2D
paradigm, D2D can 
be effectively used to offload traffic from the  cellular
infrastructure, and even to replace some microcells.

The remainder of this paper is organized as follows. After discussing
related work in Sec.~\ref{sec:relwork}, we introduce
the system under study and our main assumptions in Sec.~\ref{sec:system}. 
The network model is presented in Sec.~\ref{sec:model}.
Sec.~\ref{sec:adp} outlines the dynamic programming formulation of the
problem and our ADP solution. Results derived  
in a realistic scenario are shown in Sec.~\ref{sec:results}.
Finally, we draw our conclusions in Sec.~\ref{sec:conclusions}.

\section{Related work\label{sec:relwork}}

The deployment of a multi-tier network where cells 
use the same radio resources is highly beneficial since it allows 
traffic offloading from
macrocells to smaller cells \cite{Hossain13}.
However, such scenario imposes the adoption of ICIC techniques, 
for which a good survey
can be found in \cite{Boudreau}. 
Additionally,  eICIC
specifications in  3GPP Rel.\,10~\cite{3GPP-eICIC} foresee the use of  the Cell
Range Expansion (CRE) in LTE systems. Such technique  
consists in adding a positive range expansion bias to the pilot
downlink signal strength received from microcells  so that more users connect to 
them. Then, in order to mitigate the interference between
overlaying macro- and microcells, macrocells may periodically mute their downlink transmissions
in certain subframes 
 (called almost blank subframes - ABSs). 
By using ABSs for edge users,  microcells can significantly improve
their performance \cite{ToN13}. 
In our work, we do not focus on eICIC techniques, rather, we
take a scenario implementing them as our term of comparison.
Unlike  the above works, we assume the presence of an 
area controller that issues resource allocation and
scheduling instructions to  BSs, through high-speed optical fiber
connectivity \cite{freescale,ericsson}. Also, we assume both I2D and  D2D communication
paradigms in all cells.

How D2D communication can be integrated with 
cellular networks and the applications it can support are
investigated in~\cite{fodor-design-issues}.
This work presents a conceptual framework for the formulation
of problems such as peer discovery, scheduling and resource
allocation. 
The problem of resource allocation is also studied in~\cite{zulhasnine,noi-mascots13}, where however
 only macrocell BSs  and D2D mode are
 considered. Additionally, in~\cite{zulhasnine} 
 the D2D  pairs wishing to exchange data are given at the outset (i.e., unlike
 our work,~\cite{zulhasnine} does not address the  
endpoint association problem).
Both \cite{zulhasnine,noi-mascots13} formulate resource allocation as a mixed integer
optimization problem, which is notoriously
hard to solve, with \cite{zulhasnine} also presenting  a greedy heuristic. 
The work in~\cite{lin-arxiv} further compounds the problem by
investigating the selection of the most
suitable communication mode, still in
a single-tier scenario with D2D. There,
an analytical model is proposed, based on the assumption that the
positions of BSs and users can be modeled as a Poisson point process. Beside the different
methodology and scope of the above studies, we stress that 
our work addresses HetNets including macrocells, microcells and D2D.
While~\cite{lin-arxiv} derives  an optimal factor of
spectrum partition between cellular and D2D communication, we aim at 
determining the endpoint that should serve each user and an efficient
data scheduling, on a single radio resource basis.

\section{System scenario and assumptions\label{sec:system}}

We consider a two-tier HetNet, including LTE-based  
macrocells and microcells deployed in a urban environment. 
Each cell, either macro or micro, is controlled by a base station
(BS), which is referred to as  macroBS in the former case and as
microBS in the latter. 
Given the new, complex tasks and the ever increasing amount of traffic that the cellular
infrastructure  is expected to handle, 
we assume that BSs have optical fiber connectivity to  the core network, as envisioned 
by operators and network manufacturers \cite{freescale,ericsson}.

The coverage of a BS (either 
macroBS or  microBS), is given by the area where the received
strength of the BS pilot signal is higher than -70\,dBm  \cite{ITU}. 
A UE under the coverage of both a
macroBS and a microBS can be served by either of them. 
I2D and D2D information transfers take place in the same band and share the same frequency
spectrum, i.e., we assume in-band, underlay D2D communication. Indeed,
as shown in~\cite{lin-arxiv}, the in-band  underlay D2D
mode outperforms the overlay mode in terms of achieved throughput.
In particular, in this work we focus on the LTE downlink spectrum, although
our model can be easily extended to consider other 
frequency bands, either uplink or unlicensed spectrum portions. 
This choice is motivated by the fact that most of the mobile and web
traffic is represented by downloads
from the Internet~\cite{arnaud}. 
Additionally, based on the recent trend and standardization
activities, 
we consider network-controlled (or, equivalently, operator-controlled) D2D
communication~\cite{underlay-twc,shen-opcontrol,3gpp-r12}. 
This implies that,
not only synchronization and security issues can be easily solved, but
also UE
pairs can be efficiently scheduled so as to use cellular resources
even at high traffic load.

We focus on unicast data transfers and assume that UEs
can be served by only one endpoint  at the time.  
Considering the most popular types of terminals, we also assume UEs to
be half-duplex, i.e., they cannot transmit and receive at the same
time. In downlink direction, 
this implies that a UE receiving information from the cellular
infrastructure cannot simultaneously serve another UE. 

According to the LTE specifications~\cite{lte-book}, 
the minimum resource scheduling unit is referred to 
as a radio block (RB). One RB consists of 12 subcarriers
(each 15~kHz wide) in the frequency domain and one subframe (1-ms long) in the time domain. 
Radio resource allocation is updated every subframe by an area
controller in the core network, which assists BSs in radio resource
allocation and traffic scheduling. 
The area controller collects information on the 
channel quality from the BSs and receives content requests from the users.
Note that BSs are
oblivious to higher-layer demands, namely, user content requests. 
From the collected information, the area controller allocates
resources, i.e., it  
determines (i) which  endpoint (among the possible ones:
macroBS, microBS, or UE) should serve each user, and (ii) 
which RB(s) to employ  for such 
communication. 
Decisions taken by the area controller are issued to the BSs, which
forward them to the UEs. 

In Sec.~\ref{sec:results}  we  compare the performance of the proposed system 
to a distributed scenario reflecting today's networks, where D2D is not supported and UEs are always
served according to the proportional fairness algorithm, 
by the BS whose received signal is the strongest.

\begin{figure}
\psfrag{U1}[r][]{\footnotesize{$u_1$}}
\psfrag{U2}[r][]{\footnotesize{$u_2$}}
\psfrag{U3}[r][]{\footnotesize{$u_3$}}
\psfrag{U4}[r][]{\footnotesize{$u_4$}}
\psfrag{U5}[r][]{\footnotesize{$u_5$}}
\psfrag{U6}[r][]{\footnotesize{$u_6$}}
\psfrag{M1}[t][]{\footnotesize{$M_1$}}
\psfrag{M2}[t][]{\footnotesize{$M_2$}}
\psfrag{P1}[l][]{\footnotesize{$m_1$}}
\psfrag{P2}[l][]{\footnotesize{$m_2$}}
\psfrag{P3}[l][]{\footnotesize{$m_3$}}
\psfrag{R1}[l][]{\footnotesize{$r_1$}}
\psfrag{R2}[][]{\footnotesize{$r_1$}}
\psfrag{R3}[][]{\footnotesize{$r_2$}}
\psfrag{R4}[c][]{\footnotesize{$r_3$}}
\centering
	\includegraphics[width=0.48\textwidth]{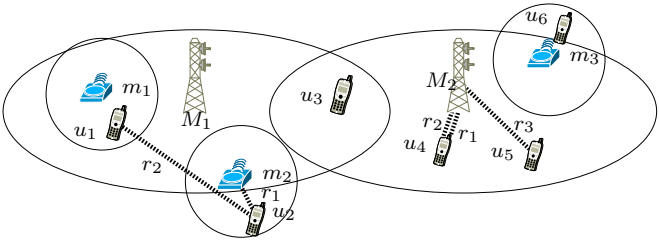}
\caption{\label{fig:scenario} An example scenario.
UEs are denoted by $u_1,\ldots,u_6$, macroBSs by $M_1,M_2$ and 
microBSs by $m_1,m_2,m_3$.
Solid lines denote coverage areas. 
Dotted lines 
correspond to RBs 
used by a pair of endpoints. 
}
\vspace{-2mm}
\end{figure}

\section{Network model\label{sec:model}}
We now build our model for the LTE-based network described in
Sec.~\ref{sec:system}.
In the following, we denote sets of elements by calligraphic-capital
letters and elements of a set by lower-case letters.
Auxiliary variables are represented by lower-case
Greek letters. 
The dependency on time appears as a superscript, while that on a
radio resource (RB), or on a content, as subscript.
The main symbols we use can be found in Table~\ref{tab:notation}.
In the text we may also refer to variables through the corresponding symbol, but
omitting their dependency on the system parameters.

\subsubsection{Base stations, users and radio resources}
We denote by $\Bc$ the set of all BSs. 
Elements in~$\Bc$ correspond to different kinds of network
infrastructure, 
namely, macro- and microBSs. 

We refer to a user equipped with a mobile terminal as, equivalently,
user or UE, and  define~$\Uc$ as the set of users in the network area.

The set  of radio resources that can be assigned to
a transmission is denoted by $\Rc$,
i.e., $r\in\Rc$ is an  RB in the downlink direction. 
Recall that RBs are assigned to transmitters every 1~ms-subframe.
We therefore divide time into a set of \emph{time steps}~$\Kc$, each assumed 
to be equal to one subframe.  
In principle, all network nodes can use any RB at the same time, though
each node uses its RBs in a time step to transmit to one other node only.
Also, a UE can be served by only one endpoint during one time step.

Endpoints of communication in our system depend on the chosen
paradigm.
Given a data flow from $e_1$ to $e_2$, $e_2$ is a downloader, while 
$e_1$ is a serving UE in D2D mode and a macroBS, or a microBS, in
I2D mode. 

\begin{table}
\caption{\label{tab:notation}List of symbols\vspace{-2mm}}
\centering
\begin{tabularx}{0.95\columnwidth}{|c|X|}
\hline 
Symbol & Description\tabularnewline
\hline 
\hline 
$\Bc$ & Set of BSs\tabularnewline
\hline 
$\Cc$ & Set of content items\tabularnewline
\hline 
$\Rc$ & Set of radio resources (RBs)\tabularnewline
\hline 
$\Uc$ & Set of users\tabularnewline
\hline 
$l_c$ & Size of content~$c$\tabularnewline
\hline 
$w_c(u)$ & Time step when user~$u$ becomes interested in content~$c$\tabularnewline
\hline 
$h_c^k(u)$ & Cumulative amount of data of content~$c$ that user~$u$
has downloaded until the beginning of time step~$k$ \tabularnewline
\hline 
$\delta_r^k(e_1,e_2)$ & Amount of data that can be sent from~$e_1$
to~$e_2$ on RB~$r$ at time step~$k$\tabularnewline
\hline 
$\chi_c^k(e_1,e_2)$ & Amount of data of content~$c$ transferred
from ~$e_1$ to~$e_2$ at time step~$k$ (over any possible RB)
\tabularnewline
\hline 
\end{tabularx}
\vspace{-3mm}
\end{table}

\subsubsection{Power and interference}
The power with which endpoint~$e_1\in\Bc\cup\Uc$ transmits to
endpoint~$e_2$ is indicated by~$P(e_1,e_2)$. For I2D (downlink)
transmissions, the value
of such  parameter  depends only on whether~$e_1$ is a macroBS or a
microBS, i.e.,~$P(e_1,e_2)=P(e_1)$~\cite{lte-book}.
Conversely, we assume that the transmit power of a serving UE in D2D communication
is subject to a closed-loop control, so that its value may depend 
on such factors as propagation conditions and
positions of either endpoints.

In addition, we define $A(e_1,e_2)$ as the signal attenuation
affecting the transmission between  endpoints~$e_1,e_2$. The attenuation
depends on both the position and the type of the endpoints (e.g., on
the height of the network node antennas). 

In all cases, from the viewpoint of our model, power and attenuation are
input values. Thus,  any assumption
about propagation conditions and power control algorithms can be
accommodated with no change to the model itself.
In particular, in order to precompute $A(e_1,e_2)$, we adopt 
the ITU urban propagation models specified in~\cite{ITU} for macro-
and microBSs, and the model in~\cite{d53-chan-models} for D2D
communication.  
It is important to stress that, by including power and attenuation figures as an input
to our model, we can obtain a remarkable level of realism, while  
keeping the complexity low.

Given the transmit power and the attenuation factor, the useful power
received at~$e_2$ from source~$e_1$ is~$P(e_1,e_2)/A(e_1,e_2)$. 
Similarly, considering a generic 
node pair $(e,u)$ communicating on the same RB where $e_2$ is
receiving, the interference suffered by~$e_2$ can be written as $P(e,u)/A(e,e_2)$.
Assuming that $e_1$ is transmitting to $e_2$ at time step $k$ on RB
$r$, the total interference experienced by $e_2$ is:  
\[I_r^k(e_2)=\sum_{\stackrel{(e,u)\,{\rm use}\,r\,{\rm at}\,k \wedge}{e:\,A(e,e_2)>0}} P(e,u)/A(e,e_2)\,,\]
while the signal to noise plus interference ratio (SINR) is yielded by 
\begin{equation}
\label{eq:sigma}
{\rm SINR}_r^k(e_1,e_2)=\frac{P(e_1,e_2)} {A(e_1,e_2) (N+I^k_r(e_2))}.
\end{equation}
We can finally map the SINR onto the  amount of data that can be transferred
from $e_1$ to $e_2$ using RB $r$ during step $k$. We indicate
this amount by~$\delta_r^k(e_1,e_2)$, and we determine its value
based on experimental measurements, as detailed later.

\subsubsection{Content and interest}
We denote by~$\Cc$ the set of content items that the users may request
(e.g., videos, ebooks, maps, web pages). 
For each content item~$c\in\Cc$, we know the size~$l_c$ and 
the maximum delay~$D_c$ with which it should  be delivered to a user 
(e.g., before the user loses interest in it).
 
For each user~$u\in\Uc$, we introduce an input parameter to the
model called {\em want-time}, $w_c(u)\in\Kc$,  defined as the time step
at which user $u$ becomes interested in content $c$.
We then indicate by~$h_c^k(u)$ the total amount of
content~$c$ that~$u$ has downloaded until the beginning of 
time step~$k$. Note that 
$0\leq h_c^k(u)\leq l_c$, and that such a quantity is non decreasing,
i.e.,~$h_c^k(u)\geq h_c^{k-1}(u), \forall k>0$.
We abuse the notation and define $h_c^k(e_1)=l_c,$ 
$\forall e_1 \in \Bc$. That is,  BSs can download
the whole content $c$ in negligible time (recall that they 
are connected to the core network through optical fibers). 
We remark that partially-downloaded content items can be transferred on 
a D2D link, though limited to the portion available at the serving UE.

Variable~$\chi_c^k(e_1,e_2)$ denotes the amount of data of content~$c$ transferred
from endpoint $e_1$ to $e_2$ during time
step $k$,
over all possible RBs. Thus, we have the following inequality:
\begin{equation}
\label{eq:chi}
\sum_{c\in\Cc}\chi_c^k(e_1,e_2)\leq\sum_{r\in\Rc}\delta_r^k(e_1,e_2).
\end{equation}
In (\ref{eq:chi}), strict inequality holds when
$e_1$ is a serving UE and the total amount of data it is caching for  $e_2$ is
smaller than what could be transferred over the link between the two nodes.

\section{A Dynamic Programming-based Approach\label{sec:adp}}

In the following, we  introduce the model we developed using the 
standard dynamic programming methodology. 
As shown by previous work \cite{zulhasnine,noi-mascots13}, the
problem of radio resource allocation in LTE-based
systems is NP-hard, even when less complex scenarios than ours 
are considered. Thus, we resort
to  approximate dynamic programming  in order to solve the model  
in realistic, large-scale  scenarios.

\subsection{The dynamic programming model\label{sec:DPmodel}}

\begin{figure*}
\psfrag{ST}[][M]{\footnotesize{$\mathbf{s}^k$}}
\psfrag{AC}[][M]{\footnotesize{$\{\mathbf{a}^k\}$}}
\psfrag{PT1}[][M]{\footnotesize{$\delta_r^k(e_1,e_2)$}}
\psfrag{AT1}[][M]{\footnotesize{$\chi_c^k(e_1,e_2)$}}
\psfrag{CON}[][M]{\footnotesize{$\mathbf{C}(\mathbf{s}^k,\mathbf{a}^k)$}}
\psfrag{VAL}[][M]{\footnotesize{$\mathbf{V}(\mathbf{s}^k,\mathbf{a}^k)$}}
\psfrag{BAC}[][M]{\footnotesize{$\min_a(\mathbf{C}+\mathbf{V})$}}
\psfrag{PT2}[][M]{\footnotesize{$\delta_r^{\star}(e_1,e_2)$}}
\psfrag{AT2}[][M]{\footnotesize{$\chi_c^{\star}(e_1,e_2)$}}
\psfrag{REQ}[][M]{\footnotesize{\hspace{-.5cm}$w_c(u)$}}
\psfrag{kpp}[][M]{\footnotesize{$k\gets k+1$}}
\centering
	\includegraphics[width=0.25\textwidth]{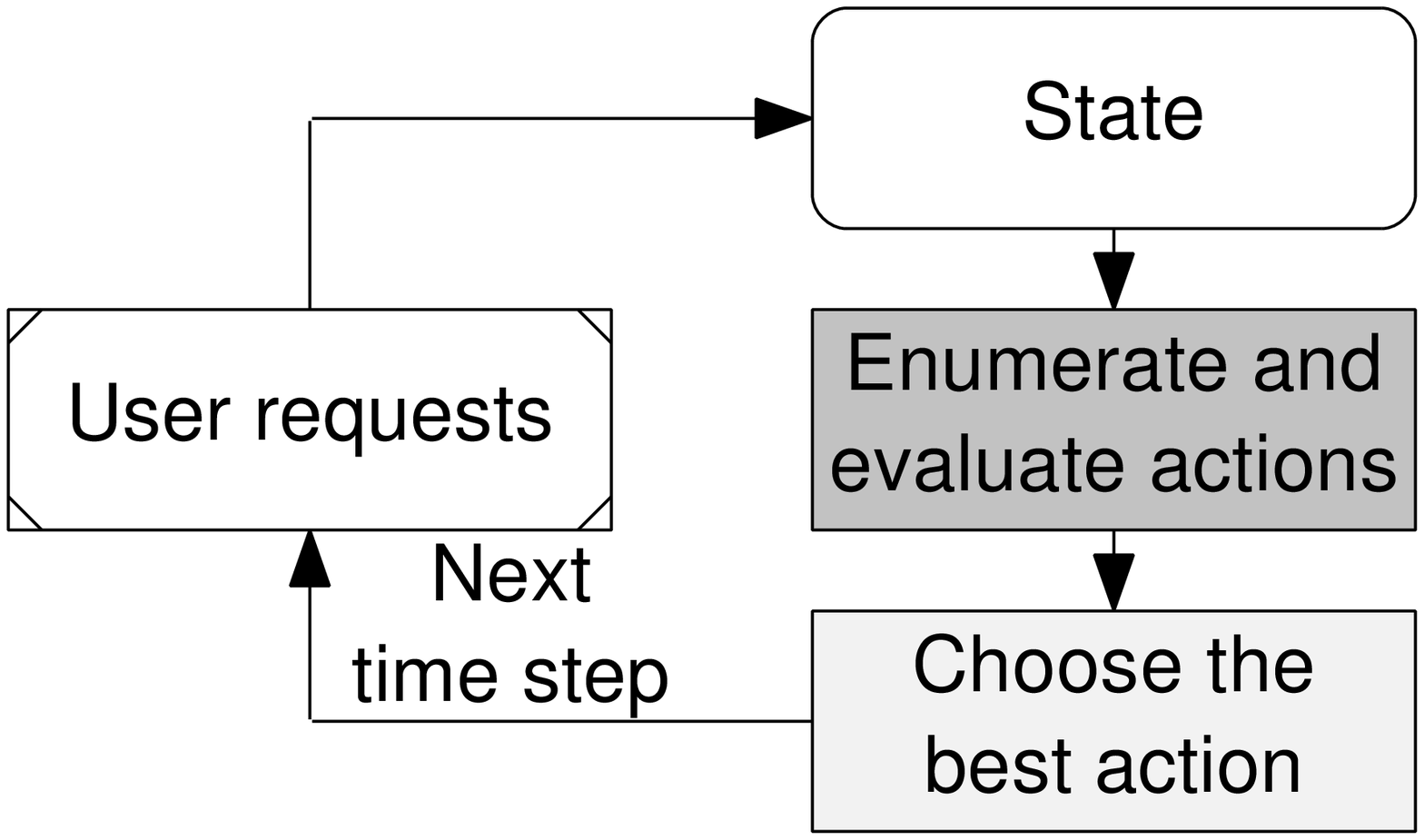}
\hspace{.2cm}
	\includegraphics[width=0.72\textwidth]{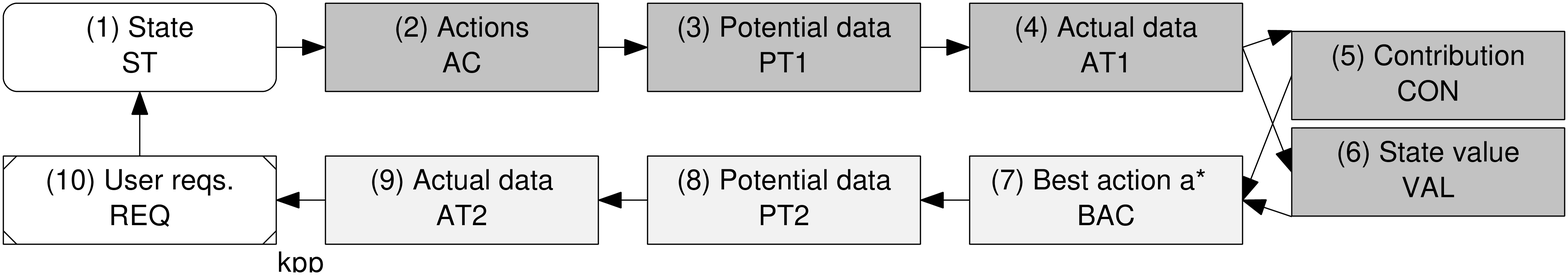}
\caption{\label{fig:dp-flow}Dynamic programming. Left: main steps involved.
Right: detailed view. Given the current state (1), the set of possible
actions can be determined (2). For each action, it can be computed 
the potential (3) and actual (4) amount of
content transferred between the pairs of endpoints. These values are used to compute the
cost (5) of an action, and to estimate the value of the state
it leads to (6).
The latter two figures are used (7) to select the best action.  
The resulting transfers (8-9), along with
the users that just became interested in a content, define the next
state.}
\vspace{-3mm}
\end{figure*}

Dynamic programming is an optimization technique based on breaking a
complex problem into simpler, typically
time-related, subproblems. Since scheduling in
LTE systems occurs every subframe, we solve the
resource allocation problem every time step $k$.  
A dynamic programming model consists of the following
elements  
(denoted by bold-face Latin letters)~\cite{adp-book}:
\begin{itemize}
\item the \emph{state variable}, $\sv^k$, which describes the state of the system 
at time~$k$; 
\item the \emph{action} set, $\Ab^k=\{\av^k\}$ i.e., all possible
  decisions that can be  taken at  time~$k$; 
\item an exogenous (and potentially stochastic) \emph{information
    process}, accounting for  information on the system becoming
  available at time $k$; 
\item the \emph{cost} of an action, $\Cb(\sv^k,\av^k)$, i.e., the
  immediate cost due to the selected action, given the current state; 
\item the \emph{value}, $\Vb(\sv^k,\av^k)$,  of ending up at a new
  state $\sv^{k+1}$, determined by the current state and action; such 
  value is   
given by the cost associated with the optimal system evolution from $\sv^{k+1}$. 
\end{itemize}
Table~\ref{tab:dp-symbols} summarizes these quantities, their meaning
in our system 
and  the symbols we use for them. Fig.~\ref{fig:dp-flow} shows how
each of them is used in the model.

\begin{table}
\caption{\label{tab:dp-symbols}Dynamic programming model\vspace{-2mm}}
\centering
\begin{tabularx}{0.95\columnwidth}{|X|X|}
\hline 
Quantity and symbol & Description\tabularnewline
\hline 
\hline 
Current state $\sv^k$ & Set of duplets, each referring to a different 
user-content pair. A duplet includes the amount of content $c$ already
downloaded by $u$,~$h_c^k(u)$, and the want-time~$w_c(u)$ if
no greater than~$k$ 
\tabularnewline\hline 
Action to take~$\av^k$ & Set of triplets indicating which pairs of
endpoints~$(e_1,e_2)$ should communicate on which RB, i.e.,~$(e_1,e_2,r)$
\tabularnewline
\hline 
Exogenous information & Want-times $w_c(u)$ 
\tabularnewline
\hline 
Cost $\Cb(\sv^k,\av^k)$ & Ratio of the amount of
content still to be retrieved by interested users to the remaining
time before the deadline for content delivery expires\tabularnewline
\hline 
Value~$\Vb(\sv^k,\av^k)$ & Total
(expected) costs due to the system future evolution 
\tabularnewline
\hline 
\end{tabularx}
\vspace{-3mm}
\end{table}

In particular, in our case the system state at generic time $k$ is 
given by the set of  duplets: $\sv^k=\{h_c^k(u),w_c(u)\}_{u,c}$. 
Each duplet refers to a different user-content pair, $u$ and $c$, and
includes  (i) the  amount~$h_c^k(u)$ of the content downloaded by the user,
and (ii) the  want-time $w_c(u)$. Clearly, at time~$k$ we only
know those want-times~$w_c(u)\leq k$. 

An action is a set of triplets, each defining which endpoint $e_1$
should serve downloader $e_2$ and using which RB $r$, i.e.,
$\av^k=\{(e_1,e_2,r)\}$. In simpler terms, an action is a realization
of resource allocation.

The dynamic programming model works as shown in
Fig.~\ref{fig:dp-flow}~(left): 
for each time step we enumerate and evaluate
the possible actions, select (and enact) the best one, and move to the
next time step. At this point, we become 
aware of which content items have been recently requested, hence we
can determine the next system state. 

Fig.~\ref{fig:dp-flow}~(right) offers a more detailed view.
The starting point is given by the current state~$\sv^k$ and the set of actions describing
the possible resource allocations (steps 1 and 2 in the figure). 
For each action, we compute
the potential $(\delta)$ and, then, the actual $(\chi)$ amount of data that
can be transferred between every pair of
endpoints (steps 3--4). 
Given the variables $\chi$, we update the total  amount of data that
each downloader $e_2$ can obtain by
the beginning of the next time step as,
\begin{equation}
\label{eq:dp-nextstate}
h^{k+1}_c(e_2) 
\gets
h^k_c(e_2)+\sum_{e_1\in\Bc\cup\Uc}\chi^k_c(e_1,e_2) 
\,.
\end{equation}

For each action~$\av^k$, we can then evaluate 
the cost $\Cb(\sv^k,\av^k)$ the system incurs if $\av^k$ is selected (step 5 in
Fig.~\ref{fig:dp-flow}~(right)).
We define such cost  as the sum over all downloaders and content  of the 
ratio of the amount of data still to be retrieved by the
downloader to the time before the  content delivery deadline  expires, i.e.,
\begin{eqnarray}
\label{eq:dp-contribution} 
\Cb(\sv^k,\av^k)\hspace{-3mm}&=&\hspace{-3mm}\sum_{c\in\Cc} \hspace{-1mm} 
\sum_{\stackrel{e_2\in\Uc \colon}{w_c(e_2)\leq k}} \hspace{-3mm} \frac{l_c-\left(h_c^k(e_2)+\sum_{e_1\in\Bc\cup\Uc}\chi_c^k(e_1,e_2)\right)}{ w_c(e_2)+D_c-k}\,\nonumber\\
\end{eqnarray}
By the above definition, a lower cost
is therefore obtained for those  allocation strategies, $\av^k$, 
assigning more resources to  downloads 
that are closer to their completion deadline. 

The value~$\Vb(\sv^k,\av^k)$  (step 6 in Fig.~\ref{fig:dp-flow}~(right)) 
is yielded by the sum of the 
costs~$\Cb(\sv^{k+1},\av^{k+1})+\Cb(\sv^{k+2},\av^{k+2})+\ldots$. 
In other words, it is the cost that will be paid in the
future, after the system has reached state~$\sv^{k+1}$. 
State values do not normally admit a closed-form expression.
In standard dynamic programming~\cite[Ch.~3]{adp-book}, they are
computed by accounting for all possible states and actions, typically leading 
to an exceedingly high complexity in non-toy scenarios. We address such
an issue in the following section.

Once $\Cb(\sv^{k},\av^{k})$ and $\Vb(\sv^k,\av^k)$ have been computed for all
actions, the action  $\av^*$  minimizing the cost  
$\Cb(\sv^{k},\av^{k})+ \Vb(\sv^k,\av^k)$ is selected (step 7 in
Fig.~\ref{fig:dp-flow}~(right)). Given  $\av^*$, 
the corresponding amount  of  transferred data can be calculated (steps
8-9). This,
along with fresh information on user requests (step 10), leads to the  
next state $\sv^{k+1}$.

Next, we detail how to compute the amount of data
$\delta_r^k(e_1,e_2)$ (Alg.~\ref{alg:compute-delta}) and
$\chi_c^k(e_1,e_2)$ (Alg.~\ref{alg:compute-chi}), taking into account 
the interference due to the spatial reuse of radio resources.
It is worth stressing that, in spite of its apparent intricacy
and high level of realism, the process we describe below has a very
low computational complexity,
namely~$O(|\Uc|)$.

\begin{algorithm}
\begin{algorithmic}[1]
\Require $\mathbf{a}^k$
\State $I_r^k(u)\gets 0, \forall u\in\Uc$, $\forall r\in\Rc$\label{line:delta-init-tau}
\ForAll {$(e_1,e_2,r)\in\mathbf{a}^k$}
\ForAll {$u\in\Uc\setminus\{e_1,e_2\}$}
\State $I_r^k(u)\gets I_r^k(u)+ \mathds{1}_{A(e_1,u)>0} P(e_1,e_2)/A(e_1,u)$ \label{line:delta-increment-tau}
\EndFor
\EndFor
\ForAll{$(e_1,e_2,r)\in\mathbf{a}^k$}
\State ${\rm SINR}_r^k(e_1,e_2)\gets \frac{P(e_1,e_2)}{A(e_1,e_2) (N+I_r^k(e_2))}$ \label{line:delta-compute-sigma}
\State $\delta_r^k(e_1,e_2)\gets \text{{\bf sinr\_to\_delta}}({\rm SINR}_r^k(e_1,e_2))$ \label{line:delta-compute-delta}
\EndFor
\State \Return $\delta_r^k(e_1,e_2)$
\end{algorithmic}
\caption{\label{alg:compute-delta}Computing the amount~$\delta$ of
  data that can be potentially transferred}
\end{algorithm}
\vspace{-5mm}
\begin{algorithm}
\begin{algorithmic}[1]
\Require $\mathbf{a}^k$, $\delta_r^k(e_1,e_2)$
\State $\chi_c^k(e_1,e_2)\gets 0$, $y_{r,c}^k(e_1,e_2) \gets 0, 
\forall c,e_1,e_2, r$
\ForAll{$(e_1, e_2, r)\in \mathbf{a}^k\colon \delta_r^k(e_1,e_2)>0$}
 \While{$\sum_{c\in\Cc\colon w_c(e_2)\leq k}y_{r,c}^k(e_1,e_2)<\delta_r^k(e_1,e_2)$} \label{line:chi-while}
  \State $c^{\star}\gets \arg \min_{c\in\Cc \colon h_c^k<l_c} w_c(e_2)$ \label{line:chi-select-content}
  \State
  {$y_{r,c^{\star}}^k(e_1,e_2)~\gets~\min~\{h_{c^{\star}}^k(e_1)-h_{c^{\star}}^k(e_2),$ \quad
    $\delta_r^k(e_1,e_2) - \sum_{c\in\Cc}y_{r,c}^k(e_1,e_2)\}$} \label{line:chi-compute-y}
 \EndWhile
 \State $\chi_{c^{\star}}^k(e_1,e_2)\gets\chi_{c^{\star}}^k(e_1,e_2)+y_{r,c^{\star}}^k(e_1,e_2)$\label{line:chi-update-chi}
\EndFor
\State \Return $\chi_c^k(e_1,e_2),y_{r,c}^k(e_1,e_2)$
\end{algorithmic}
\caption{\label{alg:compute-chi}Computing the amount~$\chi$ of data being actually transferred}
\end{algorithm}
\vspace{-3mm}

Algorithm~\ref{alg:compute-delta} 
is used in  steps 3 and 8 in Fig.~\ref{fig:dp-flow}~(right).
In line~\ref{line:delta-increment-tau}, 
we account for the fact that every active endpoint pair
may create interference at other users.
All interference values are computed within the first loop. 
The second loop computes the SINR
(line~\ref{line:delta-compute-sigma}) and maps it onto the amount of
data that can be transferred on RB $r$ during time step $k$ (line~\ref{line:delta-compute-delta}).
We perform such mapping by using the experimental values in~\cite{sinr-to-delta-measures}. 

Algorithm~\ref{alg:compute-chi}  instead refers to steps 4 and 9 in Fig.~\ref{fig:dp-flow}~(right)). 
The algorithm takes as  input the
action~$\mathbf{a}^k$ and the amount of data $\delta_r^k(e_1,e_2)$ that can be potentially
transferred as a consequence of this action (computed through Alg.~\ref{alg:compute-delta}).
Then, for each pair of active endpoints and assigned RB, it  selects which
content to transmit. This is done in
line~\ref{line:chi-select-content}, giving priority to incompletely
transferred content items 
that were requested first. Note that the conditional loop in
line~\ref{line:chi-while} reflects the fact that  data from multiple
content can be accommodated in the same RB, if needed.
In particular, in line~\ref{line:chi-compute-y}, for each item 
the data transferred on RB~$r$ is determined: 
this amount, indicated by $y_{r,c^\star}^k(e_1,e_2)$, is given by the minimum between 
the amount of data that source $e_1$ still has for downloader $e_2$
and the amount of data that can still be accommodated in the RB\footnote{The computation of this  amount assumes that content
is downloaded in order, i.e., 
from the first to the last byte. It does not hold 
for p2p applications, however 
file transfers and multimedia streaming do behave this way.}. 
Finally, the $\chi$-value is obtained by summing the $y$ values 
over all RBs (line~\ref{line:chi-update-chi}).

Notwithstanding the low complexity implied by the computation of the
$\delta$ and $\chi$ quantities, 
standard dynamic programming itself is affected by the well-known ``curse of dimensionality''~\cite{adp-book}, which makes it
impractical for all but very small scenarios. What causes such problem 
 is the exceedingly large set of possible actions and the
 aforementioned complexity in the evaluation of the future cost $\Vb$.
As an example, consider 
the set~$\Ab^k$ of possible actions that can be taken at time step $k$, which includes 
all possible sets of~$(e_1,e_2,r)$ triplets. 
There are~$|\Bc\cup\Uc||\Uc||\Rc|$ such tuples and, thus, a total
of~$2^{|\Bc\cup\Uc||\Uc||\Rc|}$ possible actions~$\av^k\in\Ab^k$.
Some of these actions can be discarded as meaningless, e.g., allocating RBs to a UE
that already completed its download.
Others, e.g., having a UE receive  from more than one endpoint
in the same time step, or receiving a content while transmitting
to another UE, 
are ruled out by  technology constraints~\cite{lte-book}.
However, the very fact that the size of~$\Ab^k$ grows exponentially
with the number of UEs, BSs and RBs
makes a standard dynamic programming model not scalable.
For a similar reason, the evaluation of $\Vb$ stemming from $\Ab^k$ is exceedingly
cumbersome. Indeed, one should consider all possible system evolutions starting from the current
state, by selecting at each future time step the optimal action.
Thus, we  resort to ADP and propose the algorithms below so as to
efficiently 
generate and rank actions, hence finding a solution with  low
computational complexity.

\subsection{The ADP solution\label{ssec:adp-actions}}

Recall that the immediate cost $\Cb$ of each action can be evaluated
with  very low complexity, thanks to  
 Algs.~\ref{alg:compute-delta} and~\ref{alg:compute-chi}. 
Thus, in order to  ensure scalability, it is
sufficient to act along two directions:
(i) making the number of actions to be evaluated at each time step
smaller and independent of the number
of UEs and BSs, and (ii) reducing the complexity of 
evaluating the future cost $\Vb$ of an  action. 
Of course, it is not possible to achieve such a result while
keeping the optimality guarantee. 
However, such an approach has been shown to be very
effective~\cite[Ch.~1]{adp-book}, as also confirmed by our performance
evaluation in Sec.~\ref{sec:results}. 

Below, we describe how we tackle the two issues. 

\subsubsection{Reducing the action space}
We define an auxiliary action space~$\tilde{\Ab}^k$, whose size is much
smaller than the original action space~$\Ab^k$ and, more importantly, does not grow with the
number of UEs or BSs. Then, we show
a deterministic (and computationally efficient) way to map an action~$\tilde{\av}^k\in\tilde{\Ab}^k$
of the auxiliary action space into an action~$\av^k\in\Ab^k$. 
It follows that the actions we evaluate (steps 5--7 in
Fig.~\ref{fig:dp-flow}~(right))  
are only those $\av^k\in\Ab^k$
that have a correspondence in $\tilde{\Ab}^k$.

To determine the auxiliary action space, we proceed as follows.
We ask ourselves what kind of choice has the highest relevance 
in a system such as ours. The most significant one is to rank 
transfer paradigms, i.e., using macroBSs, microBSs or D2D -- 
and test which combination of them yields the highest throughput
and carries the least interference.
We thus represent the ``importance'' of each paradigm
by a triplet of real values~$\alpha_M,\alpha_m,\alpha_u\in [0,1]$.
These values indicate which endpoints should be preferably used, 
as shown in Alg.~\ref{alg:choose-endpoints}, and each triplet
represents an auxiliary action $\tilde{\av}^k$. For the set of auxiliary
actions to be manageable, we need to discretize each value in the $\alpha$ triplet.
The set~$\tilde{\Ab}^k$ is thus finite and we can control its size by choosing 
the granularity of each $\alpha$. This is our tuning knob for
scalability purposes.

Algorithm~\ref{alg:choose-endpoints} takes as input an action~$\tilde{\av_k}$
and maps it onto an action~$\av_k$ (line~\ref{line:ce-return}).
Its logic is  straightforward: we serve downloaders, 
starting from the neediest ones, selecting the most
effective endpoint.

More specifically, in line~\ref{line:ce-downloaders}, we identify the set~$\Dc\subseteq\Uc$ of downloaders, i.e., users
with an incomplete download. This set is sorted
(line~\ref{line:ce-sort}) by the want-time~$w_c(u)$, so that
users that required the content first are given higher priority. Then, for each downloader~$u\in D$,
we loop over the potential source endpoints~$e$ and RBs~$r$ that~$e$ may use to transmit to~$u$
(line~\ref{line:ce-forall-ec}). 
For each~$(e,r)$ pair, we compute a score~$\sigma$, which is initialized
(line~\ref{line:ce-def-sigma}) to the amount of data (computed by
Alg.~\ref{alg:compute-chi}) that~$u$ may download from~$e$.
Lines~\ref{line:ce-alpha-m}-\ref{line:ce-alpha-u} play out the prioritization role
of the~$\alpha_M,\alpha_m,\alpha_u$ coefficients as follows. We weight the $\sigma$ scores by 
multiplying them by the~$\alpha$-coefficient corresponding to the type
of endpoint~$e$. For convenience, we spell out the subsets including macro- and microBSs 
as $\Bc_M$ and $\Bc_m$, respectively.
As an example, the~$\alpha$-coefficients give us leverage to encourage D2D transfers by setting a high value
for~$\alpha_u$, or to limit the usage of macroBSs to users that have no other means to be served by setting
a low value  for~$\alpha_M$.
In line~\ref{line:ce-select-best}, 
we select the endpoint corresponding to the highest sum of 
scores over all possibles RBs. 
Notice that by selecting only one endpoint in line~\ref{line:ce-select-best},
we honor the technology constraint by which each user can download data from at most one source in a given
time step.
In the following line, 
we assign to the endpoint pair $(e^\star,u)$ the RB that maximizes
their $\sigma$ score. 
However, before including the new triplet $(e^\star,u,r^\star)$ in the
allocation yielded by $\av^k$, we check whether   
the total amount of data transferred in the network increases or not
(lines~\ref{line:check-gain-B}--\ref{line:check-gain-E}). 
While verifying that, we resort again to Algs.~\ref{alg:compute-delta}
and \ref{alg:compute-chi} 
to compute the $\delta$ and $y$ values.  If the
amount of data grows,  
the triplet is added to action~$\av^k$
(line~\ref{line:ce-add-a}).

In conclusion, we stress that the size of the auxiliary action
space~$\tilde{\Ab}$ is small and it is independent of 
the number of UEs and BSs. We thus achieved our scalability goal.

\begin{algorithm}[h]
\begin{algorithmic}[1]
\Require $\tilde{\av}^k=\left (\alpha_M,\alpha_m,\alpha_u\right)$ \label{line:ce-input}
\State $\Dc\gets \{u\in\Uc~{\rm s.t.}~\exists c\in\Cc \colon w_c(u)<k \wedge h^k_c(u)<l_c\}$ \label{line:ce-downloaders}
\State $\text{{\bf sort} } \Dc \text{ {\bf by} } w_c(u)$ \label{line:ce-sort}
\ForAll{$u\in \Dc$}
 \ForAll{$e,r$} \label{line:ce-forall-ec}
\State {\bf compute} $y_{r,c}^k(e,u), \forall c\in\Cc$ (Alg.~\ref{alg:compute-chi}) 
  \State $\sigma(e,r)\gets\sum_{c\in\Cc}y_{r,c}^k(e,u)$ \label{line:ce-def-sigma}
   \State {\bf if} {$e\in\Bc_M$} {\bf then}
    $\sigma\gets\sigma\cdot\alpha_M$ \label{line:ce-alpha-m}
  \State {\bf if} {$e\in\Bc_m$} {\bf then}
    $\sigma\gets\sigma\cdot\alpha_m$ \label{line:ce-alpha-p} 
  \State {\bf if} {$e\in\Uc$}  {\bf then}
  $\sigma\gets\sigma\cdot\alpha_u$ \label{line:ce-alpha-u}
 \EndFor
 \State $e^{\star}\gets\arg\max_e\sum_r\sigma(e,r)$ \label{line:ce-select-best}
 \State $r^{\star}\gets\arg\max_r\sigma(e^{\star},r)$ \label{line:r-select-best}
 \State $t_{curr}\gets 0$, $ t_{new}\gets 0$ 
 \ForAll {$(e_1,e_2,\rho) \in \av^k$ and $c\in\Cc$} \label{line:check-gain-B}
   \State {\bf compute} $\delta_\rho^k(e_1,e_2)$  {\bf and}
   $y_{\rho,c}^k(e_1,e_2)$ (Algs.~\ref{alg:compute-delta}-\ref{alg:compute-chi}) \label{line:comp-d-y-curr}
   \State $t_{curr} \gets t_{curr}+ y_{\rho,c}^k(e_1,e_2)$ \label{line:comp-t-curr}
 \EndFor
 \ForAll {$(e_1,e_2,\rho) \in \av^k\cup (e^{\star},u,r^{\star})$ and $c\in\Cc$}
   \State {\bf compute} $\delta_\rho^k(e_1,e_2)$  {\bf and}
   $y_{\rho,c}^k(e_1,e_2)$ (Algs.~\ref{alg:compute-delta}-\ref{alg:compute-chi}) \label{line:comp-d-y-new}
   \State $t_{new} \gets t_{new} +y_{\rho,c}^k(e_1,e_2)$\label{line:comp-t-new}
 \EndFor
 \If{$t_{new} > t_{curr} $} \label{line:cfr-convenience} \label{line:check-gain-E}
   \State $\av^k\gets\av\cup (e^{\star},u,r^{\star})$ \label{line:ce-add-a}
 \EndIf
\EndFor
\State \Return $\av^k$ \label{line:ce-return}
\end{algorithmic}
\caption{\label{alg:choose-endpoints}Mapping $\alpha$-triplets 
  into actions}
\end{algorithm}
\vspace{-3mm}

\subsubsection{Evaluating the state values\label{ssec:adp-states}}
To evaluate an action, it is important to compute the value of
the state $\sv^{k+1}$ the action leads to.  
As already stated, 
the value of a state corresponds to the
sum of the costs we will pay due
to future actions, if these are chosen optimally. 
Clearly, if we set $\Vb(\sv^k,\av^k)=0$ for all actions, i.e., we
select the action that seems more profitable at the
current step, we end up adopting a greedy strategy. 
However, in network scenarios where D2D is allowed, a 
more balanced approach  accounting for future actions may be
of particular relevance. Indeed, 
transmitting to some users at a faster pace, so that they 
can act as serving UEs later, may benefit the whole network.

It follows that we need to compute 
the value function~$\Vb$ accurately enough,
while keeping the complexity low.
To do so, we resort to the methodology typically used in ADP. Such methodology
\cite[Ch.~9]{adp-book} implies that, at each step $k$, we fix the
sequence of future actions, starting from state $\sv^{k+1}$.
We apply this procedure to our problem as described in Alg.~\ref{alg:estimate-state}.

The algorithm  takes as input: (i) the current state~$\sv^{k}$ and the
current action
to be evaluated~${\av}^{k}$ (i.e., the two elements determining  next step
$\sv^{k+1}$), and (ii) the future actions that we expect will be taken. 
In order to compute the latter, we start by assuming that the 
conditions experienced by a user do not change during its download time.  
This is a fair assumption since, as shown by our numerical results, users
complete their download in few seconds ($\leq5$~s), hence the movement
of pedestrian users during content download is negligible. 
Also, note that the procedure for computing the value function $\Vb$
is repeated at every time step $k$. 
We feed such information to a Markov chain-based machine learning
model, so as to  
 compute actions $\{{\av}^{k+1},\ldots,{\av}^K\}$ 
\cite[Ch.~9]{adp-book}. 

\begin{algorithm}
\begin{algorithmic}[1]
\Require $\sv^k$, ${\av}^{k}$, $\{{\av}^{k+1},\ldots,{\av}^K\}$ \label{line:es-input} 
\State $v\gets 0$
\For{$q=k+1\to K$} \label{line:es-loop-k}
\ForAll{$(e_1,e_2,r)\in {\av}^{q}$}
\State {\bf compute} $\delta_r^q(e_1,e_2)$ {\bf using} Alg.~\ref{alg:compute-delta} \label{line:compute-delta} 
\EndFor
\ForAll{$(e_1,e_2) \colon \exists\, \delta_r^q(e_1,e_2)>0$} 
\ForAll{$c\in\Cc \colon w_c(u)\leq k \wedge h_c^q(u)<l_c$}
\State{{\bf compute}~$\chi_c^q(e_1,e_2)$~{\bf using}~Alg.~\ref{alg:compute-chi}}\label{line:compute-chi}
\State$\hat{h}_c^{q+1}(e_2)\gets \hat{h}_c^{q}(e_2)+\chi^{q}_c(e_1,e_2)$ \label{line:es-loop-update}
\EndFor
\EndFor
\State {\bf compute} $\Cb(\sv^q,\av^q)$ 
\State $v\gets v+ \Cb(\sv^q,\av^q)$ \label{line:es-loop-incr}
\EndFor
\State \Return ${\Vb}(\sv^{k},\av^{k})=v$ \label{line:es-output}
\end{algorithmic}
\caption{\label{alg:estimate-state}Estimating the value of a state}
\end{algorithm}
\vspace{-3mm}

Next, we exploit the estimated information on the system 
to compute, at each future time step $q>k$, 
the $\delta$ and $\chi$ values for
each communication foreseen by action ${\av}^{q}$ 
(lines~\ref{line:compute-delta} and \ref{line:compute-chi}). 
To this end, we resort to the low-complexity algorithms presented in
Sec.~\ref{sec:DPmodel}, which account for interference. 

In line~\ref{line:es-loop-update}, for each step $q>k$, 
given the previous state and the $\chi$ values, we apply (\ref{eq:dp-nextstate}) and update the
amount of data of content $c$, $h_c^q(e_2)$, that each downloader $e_2$ can retrieve
until step $q$. 
Then, we use the quantities $\chi$ and $h$ to evaluate the cost
of action $\av^{q}$.  Note that we cannot predict future user
requests, however,  
due to the short time span before a user download completion, their number is
limited. Additionally,  their deadline will be further away in
time\footnote{Recall that Alg.~\ref{alg:estimate-state} is repeated at every time step
$k$.}, hence their impact is minimal (see (\ref{eq:dp-contribution})).
At last, ${\Vb}(\sv^{k},\av^{k})$ is calculated by summing all
future cost contributions (line~\ref{line:es-output}).

\begin{figure*}[hbt]
\subfigure[]{\includegraphics[width=0.26\textwidth]{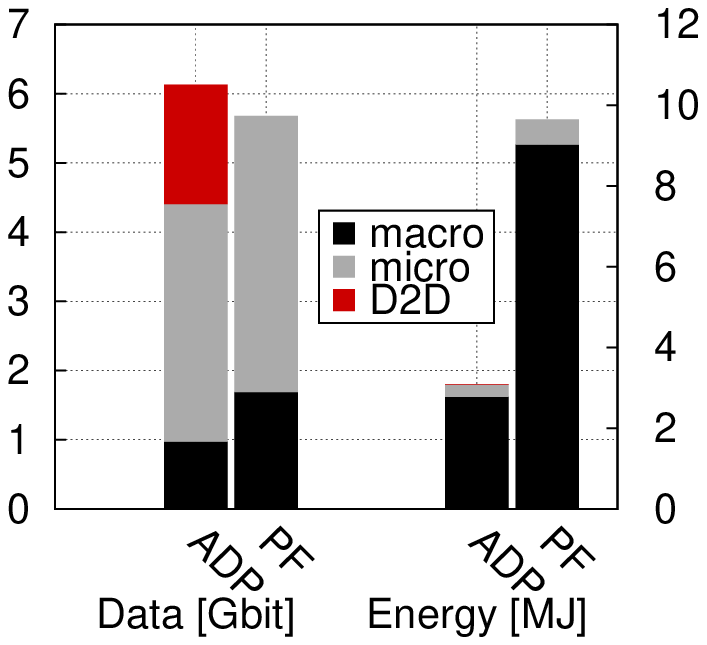}}\hspace{-6mm}
\subfigure[]{\includegraphics[width=0.25\textwidth]{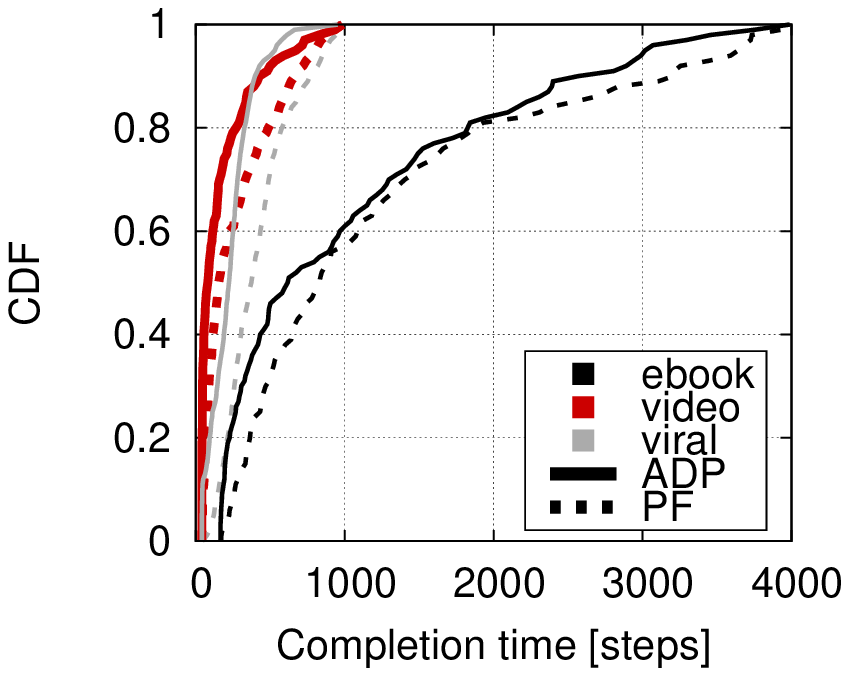}}\hspace{-2mm}
\subfigure[]{\includegraphics[width=0.25\textwidth]{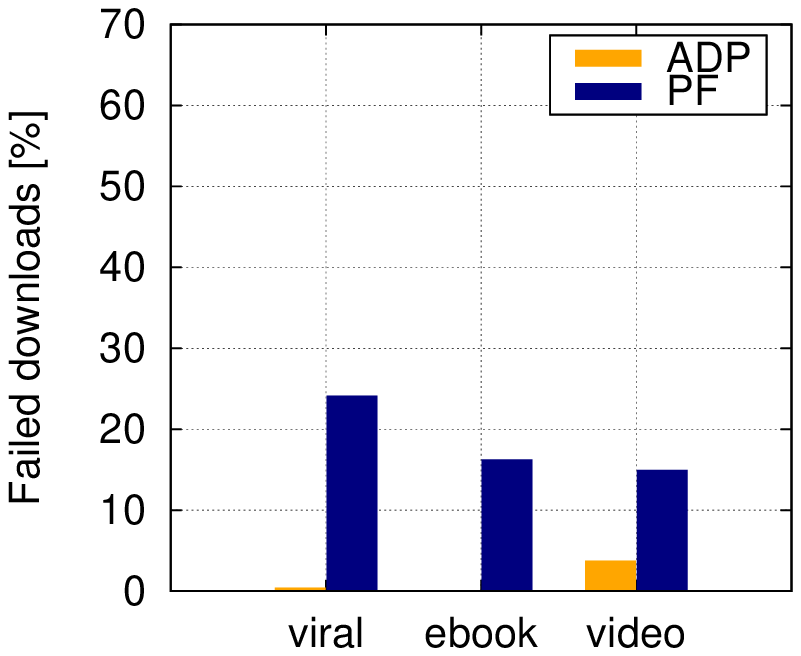}}\hspace{-2mm}
\subfigure[]{\includegraphics[width=0.25\textwidth]{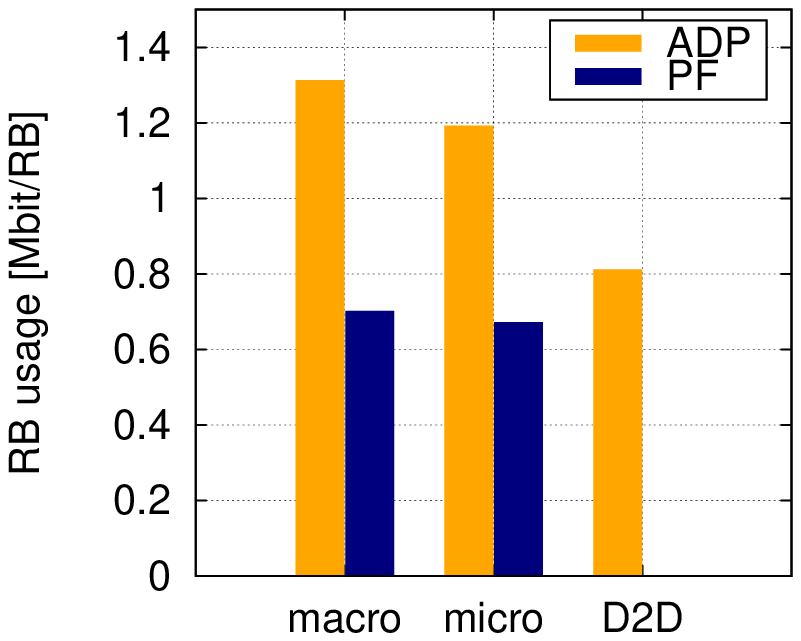}}
\vspace*{-2mm}
\caption{\label{fig:neu-joint}ADP vs. PF: total amount of
  transferred data and consumed energy (a); CDF of the completion time (b),
   failed downloads (c), RB usage (d).} 
\vspace{-3mm}
\end{figure*}
\begin{figure*}
\subfigure[ADP]{
	\includegraphics[width=0.25\textwidth]{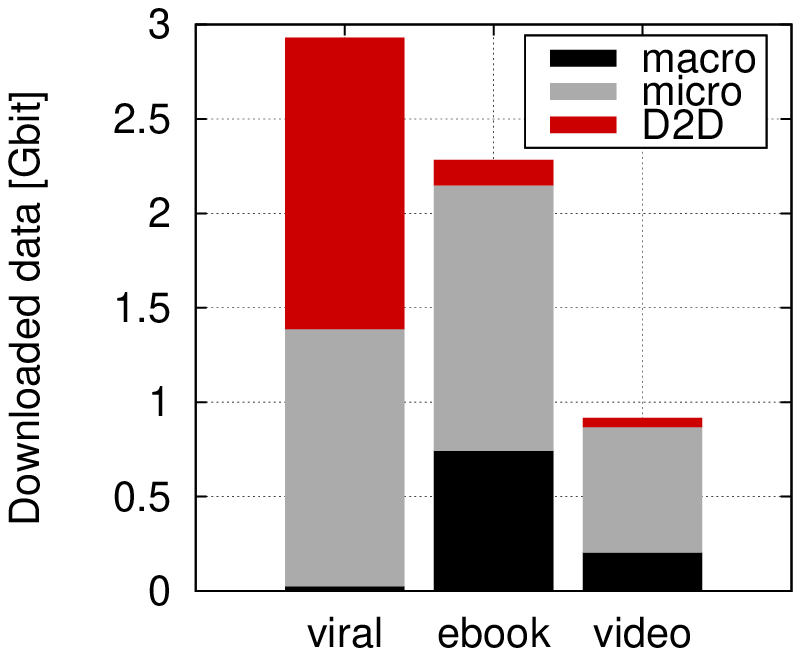}}\hspace{-3mm}
\vspace*{-1mm}
\subfigure[PF]{	\includegraphics[width=0.25\textwidth]{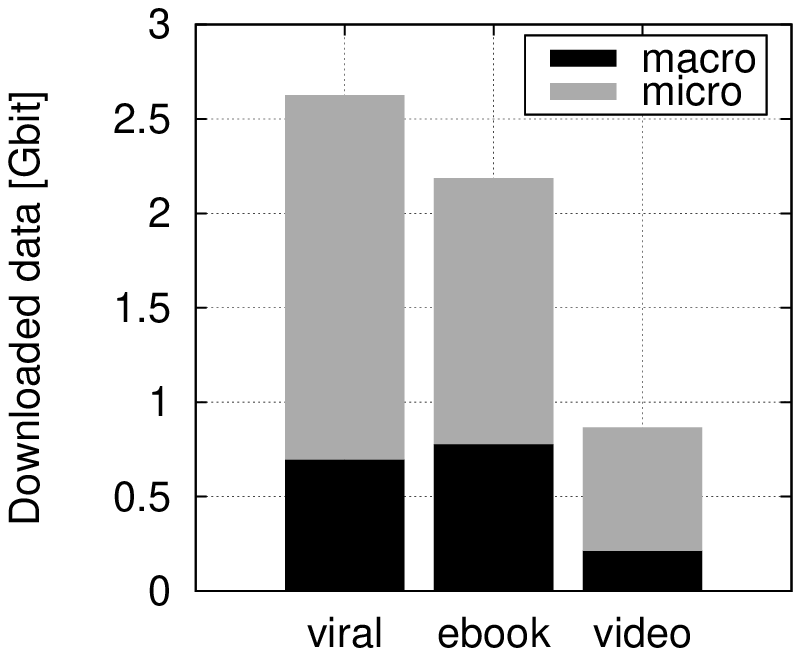}}\hspace{-3mm}
\vspace*{-1mm}
\subfigure[ADP]{	\includegraphics[width=0.25\textwidth]{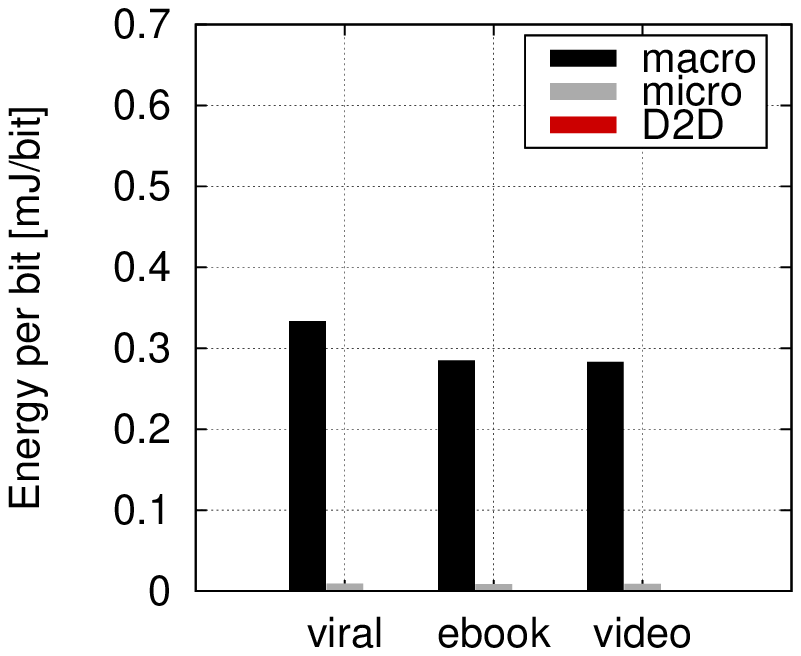}}\hspace{-3mm}
\vspace*{-1mm}
\subfigure[PF]{	\includegraphics[width=0.25\textwidth]{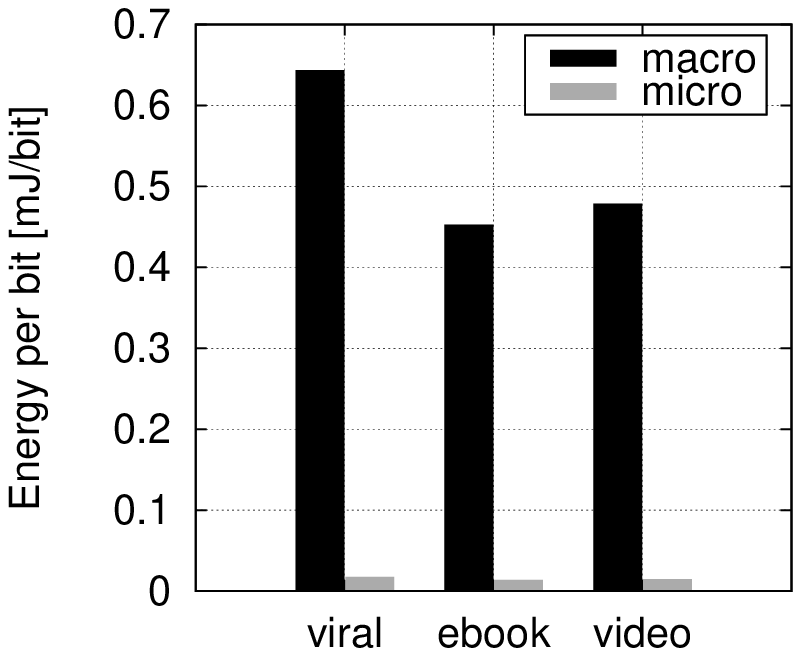}}
\vspace*{-1mm}
\caption{\label{fig:neu-sep}ADP vs. PF: breakdown of the amount of transferred
  content (a,b), and of the energy consumption per bit of
  transmitted data  (c,d).} 
\vspace{-3mm}
\end{figure*}
\begin{figure*}
\subfigure[]{	\includegraphics[width=0.26\textwidth]{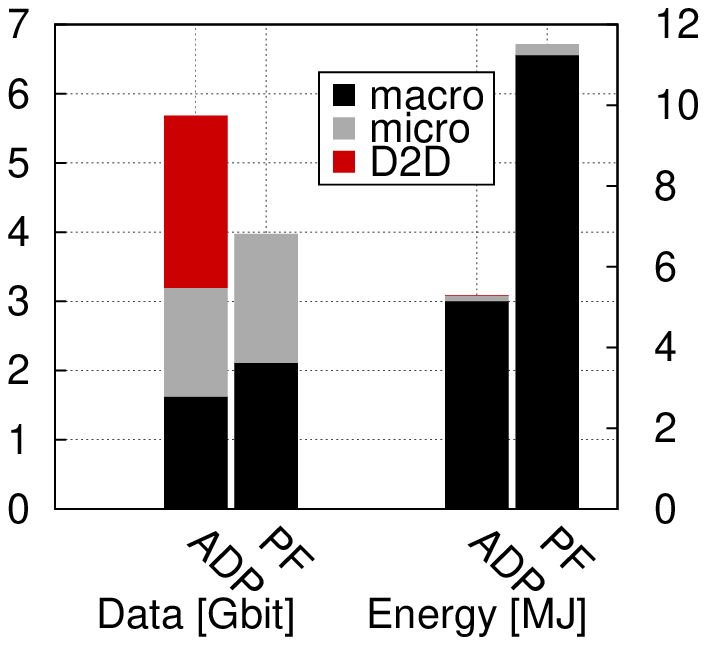}}\hspace{-6mm}
\vspace*{-1mm}
\subfigure[]{	\includegraphics[width=0.25\textwidth]{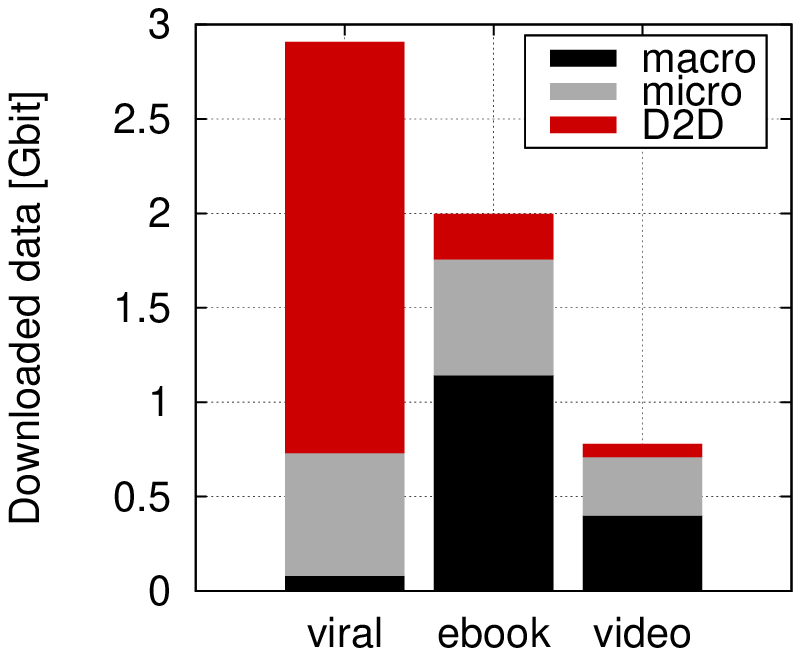}}\hspace{-2mm}
\vspace*{-1mm}
\subfigure[]{	\includegraphics[width=0.25\textwidth]{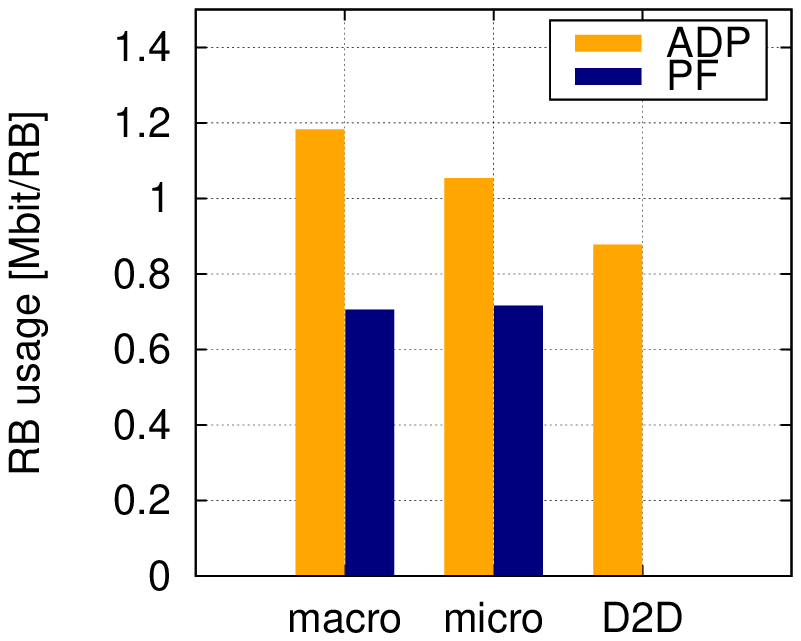}}\hspace{-2mm}
\vspace*{-1mm}
\subfigure[]{	\includegraphics[width=0.25\textwidth]{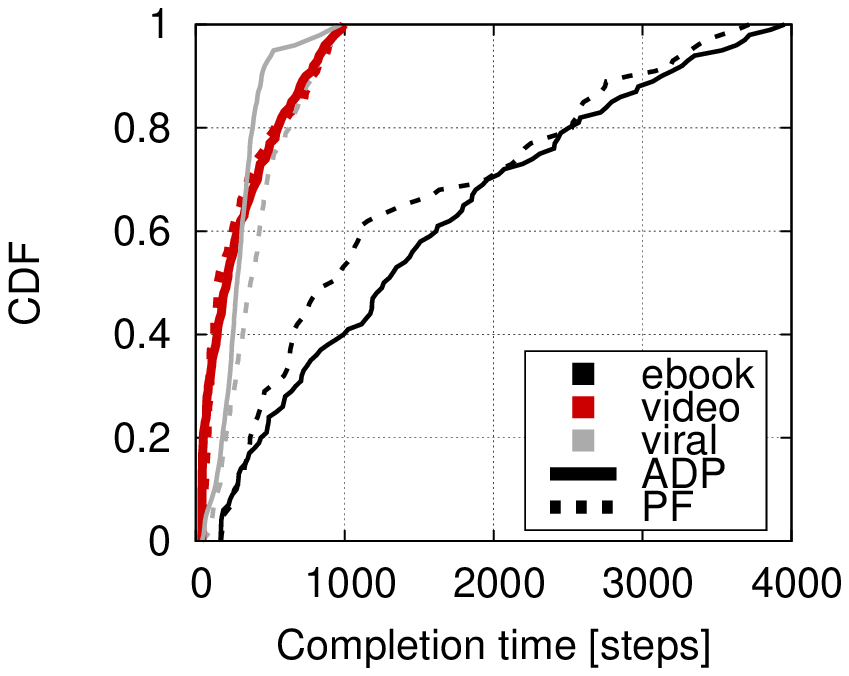}}
\vspace*{-1mm}
\caption{\label{fig:fewerbs-joint}Halving the number of microcells:
  amount of transferred data and consumed energy (a); amount of transferred 
  data by ADP (b); average RB usage (c); CDF of the
  completion time (d).}
\vspace{-3mm}
\end{figure*}

\subsection{Solution complexity\label{ssec:complexity}}

Recall that our goal is to design
a low-complexity solution.
This requirement is indeed met.
With reference to Fig.~\ref{fig:dp-flow}~(right), and assuming that 
the dominant factor is the number of users, the complexity is 
as follows. Step (2), $O(2^{|\Uc|})$ with plain dynamic programming, 
which reduces to $O(|\Uc|)$ using Alg.~\ref{alg:choose-endpoints}. 
Steps (3) and (4), linear with~$O(|\Uc|)$. Step (5),~$O(1)$. Step (6),
$O(|\Ab|^k)$ with plain dynamic programming, which reduces to 
$O(|\Uc|)$ with Alg.~\ref{alg:estimate-state}.

\section{Results\label{sec:results}}

We evaluate our solution in the two-tier scenario that is typically   
used within 3GPP for LTE network evaluation~\cite{scenario}. 
The scenario comprises  a service network area of 12.34~km$^2$, covered by 
57 macrocells and, unless otherwise specified,   228 microcells.
Macrocells are controlled by 19 three-sector BSs; 
the macroBSs inter-site distance is set to 500~m. MicroBSs are 
deployed over the network area, so that there are 4 
non-overlapping microcells per macrocell.  
A total of 3420 users are present in the area. In particular, in order
to have a higher user density where microcells are deployed,  
10 users are uniformly distributed within 50~m from each
microBS. The rest of the users are uniformly distributed over the remaining
network area.
Users move according to the cave-man model~\cite{cave-model}, with average speed of
1\,m/s. 
According to current specifications~\cite{ITU,earth}, we assume the following
pairs of values for power and antenna height: (43~dBm, 25~m) for macroBSs, 
(30~dBm, 10~m) for microBSs, and (23~dBm, 1.5~m) for UEs.
All network nodes operate over a 10~MHz band at 2.6~GHz, thus $|\Rc|=50$~RBs. 
As already mentioned, the signal propagation for I2D is modelled according to 
ITU specifications for  urban environment \cite{ITU} and for D2D
according to the specifications in \cite{d53-chan-models},  
while the SINR is mapped onto per-RB throughput values using the 
 experimental measurements in~\cite{sinr-to-delta-measures}. 
The energy consumption of the network nodes is instead computed 
according to~\cite{earth}.

Users may require content from a set of 21 different items,
belonging to three categories: ebooks, videos, or viral content; 
their characteristics and intervals between user requests
are summarized in Table~\ref{tab:contents}. We highlight
that video and viral items have stricter constraints on delivery time.
Additionally, the viral item is modeled as being in high demand
to mimic content becoming suddenly popular through social networks (the
so-called ``flash-crowd'' phenomenon).

\begin{table}[tb]
\caption{Content types\label{tab:contents}\vspace{-2mm}}
\centering
\begin{tabularx}{0.95\columnwidth}{|X|c|c|c|}
\hline 
Feature & eBook & Video & Viral\tabularnewline
\hline 
\hline 
No.\,of items & 10 & 10 & 1\tabularnewline
\hline 
Size [Mbit] & 12 & 3 & 3\tabularnewline
\hline 
Deadline [steps] & 4000 & 1000 & 1000\tabularnewline
\hline 
Request interval [steps] & 1--1000 & 1--1000 & 41--60\tabularnewline
\hline 
\end{tabularx}
\vspace{-3mm}
\end{table}

While applying our ADP approach, we consider that the values of the
$\alpha_M,\alpha_m,\alpha_u$ parameters, 
are discretized as $\{0.1,0.2,\dots,1\}$. 
Additional experiments with values exhibiting finer granularity
have  shown  negligible  improvement.
We compare our approach against a system implementing 
 the 3GPP eICIC with a microcell bias of 15\,dB and the ABS
 model where macroBSs are silent in 1 out of every 2 subframes~\cite{theory-practice}. 
In the latter, D2D mode is not supported and UEs connect to the BS
 from which they receive the strongest pilot signal. At the BSs,
 traffic is scheduled according to the proportional-fairness (PF) algorithm,
which is standard in today's LTE networks~\cite{lte-book}. In the
following, we will refer to this benchmark scenario as PF.  

The first comparison between ADP and PF is presented in
Fig.~\ref{fig:neu-joint}. Colors are used to 
differentiate among the possible endpoints (black for macroBSs,
gray for microBSs and red for UEs) and between ADP (orange) and PF (blue).
In particular, Fig.~\ref{fig:neu-joint}(a) shows that ADP allows
the transfer of more data than the state-of-the-art, 
while using a much smaller amount of energy.
Such a gain is due to the lower usage of macrocells (characterized by
very high transmit power), in favour of 
microcells and D2D. Note that the energy consumption due to D2D mode is
negligible and can be barely seen in the plot. Also, under both ADP and
PF, transmissions from microBSs are more efficient than those from
macroBSs, as the former carry a higher amount of data at a much
lower  energy cost.

Fig.~\ref{fig:neu-joint}(b) depicts the completion time of successful downloads, 
for the different content categories (denoted by a different
colors). A download is successful if it can be completed by the
corresponding deadline. First, note that, since video and viral
content have tighter deadlines, they are characterized by
better performance than ebooks. Indeed,  our cost $\Cb$
in~(\ref{eq:dp-contribution}) accounts for
content deadlines, giving higher priority to those downloads
that are closer to their completion deadline. Comparing ADP
(solid lines) to PF (dotted lines), we observe that our approach can better
meet the time requirements of content with strict deadlines (video
and viral), while guaranteeing similar delays for ebooks.

Results in Fig.~\ref{fig:neu-joint}(c) confirm the above
observation: ADP can dramatically reduce the number of failed downloads
with respect to PF.
The only content type for which ADP is unable to deliver some items is
video. This due to the fact that, in the traffic scenario under study, 
video has a quite strict deadline,
and it cannot significantly benefit from the D2D mode as users
typically ask for different items.

Finally, Fig.~\ref{fig:neu-joint}(d) highlights the improvement in ADP
usage of radio resources compared to PF. Observe that, on average, ADP can
transmit a higher amount of data per RB, as our interference-aware
scheduling assigns endpoints and radio resources far more efficiently
than the PF-based system. In other words, ADP scheduling yields higher
values of SINR, hence of data rates per RB. 
This is also underlined by the average number of times an RB is reused
in the whole network, whose value normalized to the network area  
is about 1.58 under ADP and 2.3 under PF. 
The higher value recorded under PF may at first be surprising, given that ADP 
allows D2D communication to reuse RBs too. However,  
such result further underscores the
inefficiency of PF in handling interference: it needs to reuse more RBs
in order to keep up with traffic demand.  
At last, looking at different types of endpoints, we note that RBs
assigned to macro- and microBS by PF are characterized by similar 
data rates, in spite of the lower  power irradiated by microBS.  Such behavior is
due to the use of ABS, which mutes macrocells when microcells serve
far-away UEs, and to the short distance 
between microBSs and the other UEs. 
Conversely, when ADP assigns RBs to microBSs, the difference in data rates
between macrocells and microcells flares up. 
Indeed, D2D communication causes
additional interference to UEs served by
the cellular infrastructure, which is more significant for  microBSs 
since they transmit at a lower power level.   
As for D2D mode, it exhibits slightly worse performance than 
I2D communication. This was expected, since serving UEs transmit at very low
power, hence D2D communication is more prone to interference (hence 
lower SINR and data rate).

Fig.~\ref{fig:neu-sep} presents the breakdown
of delivered data under ADP and PF, on a per-content type basis.
In spite of the lower transmission quality, D2D appears to play
a crucial role in the delivery of
viral content, as shown by Fig.~\ref{fig:neu-sep}(a).
Indeed, in case of a peak of social content demand, 
it is  likely that a downloader finds a serving UE
within its radio range. Thus, D2D can 
be effectively used to offload traffic from the  cellular
infrastructure. On the contrary, PF has to relay on macro- and
microBSs only (see Fig.~\ref{fig:neu-sep}(b)). 
As a consequence, along with a better exploitation of
radio resources, ADP requires a much lower energy 
per transferred data compared to PF, 
as evident from Figs.~\ref{fig:neu-sep}(c) and (d).
In particular, the ADP plot in Fig.~\ref{fig:neu-sep}(c) underscores that the energy consumption due
to D2D communication, normalized to the amount of downloaded data, is
negligible, thus confirming that D2D mode is a very convenient way to
spread social content.

In the scenario above, we now halve the number
of microcells from 228 to 114, i.e., 2 microcells per macrocell.
The most noticeable effect is that, with ADP, D2D communication steps up
to compensate for the missing microBSs, as shown in
Fig.~\ref{fig:fewerbs-joint}(a). Instead, PF falls short of providing the same
throughput as before. Indeed, comparing to 
Fig.~\ref{fig:neu-joint}(a), ADP exhibits a mere 8\% drop in transferred data, with respect
to 30\% for PF. Energy consumption increases for both approaches, though ADP still 
retains a clear edge. A breakdown of per-content data downloaded by ADP 
(Fig.~\ref{fig:neu-joint}(b)) shows that
D2D is even more dominant (by a 32\% increase) in viral transfers. 
In Fig.~\ref{fig:fewerbs-joint}(c), spectrum usage is less effective
with the increase in D2D communication: the surging number of D2D links
interferes more with macroBSs and the remaining microBSs. 
Those D2D links whose coverage overlaps
one of the missing microcells instead see their amount of transferred bits per RB increase.
In Fig.~\ref{fig:fewerbs-joint}(d), viral content relying more on D2D  
shows the same completion times as before, while video and ebooks
experience higher delays. 
The latter phenomenon is a consequence of the lower number of microcells.
Also, ADP tends to favour content with stricter time constraints 
(viral and video), at the expense of ebooks.
For reasons of space, we omit plots comparing other metrics, which however confirm 
the above observations.

\section{Conclusions\label{sec:conclusions}}
We considered a 2-tier, LTE-based
network, supporting D2D communication. We devised 
a solution to the problem of
selecting which endpoint should serve a user, and the radio
resources to allocate for such communication.  
In particular, we  presented approximate dynamic programming  algorithms to
generate and rank possible resource allocation decisions. In this way,
we obtained a low-complexity solution 
that can deal with realistic, large-scale scenarios.
Our results show the good performance of our solution, as well as
the conditions
under which D2D communication is more effective. 
Furthermore, we highlight that D2D mode can be a valid, low-cost
alternative to microcells in supporting traffic with little energy consumption.

\end{document}